\documentclass{article}
\usepackage{hyperref}
\usepackage{spconf,amsmath,graphicx,balance}

\usepackage{mwe} % to get dummy images
\usepackage{multirow}

\title{CryoNuSeg: A Dataset for Nuclei Instance Segmentation of Cryosectioned H\&E-Stained Histological Images}

\name{%
	\begin{tabular}{@{}c@{}}
		Amirreza Mahbod$^{\star}$, 
		Gerald Schaefer$^{\dagger}$, 
		Benjamin Bancher$^{\ddagger}$,
		Christine L\"{o}w$^{\star}$,\\
		Georg Dorffner$^{\ddagger}$, 
		Rupert Ecker$^{\mathsection}$, 
		Isabella Ellinger$^{\star}$
\end{tabular}}

\address{$^{\star}$ Institute for Pathophysiology and Allergy Research, Medical University of Vienna, Austria \\
$^{\dagger}$Department of Computer Science, Loughborough University, U.K.\\
$^{\ddagger}$Section for Artificial Intelligence and Decision Support, Medical University of Vienna, Austria\\
$^{\mathsection}$ Department of Research and Development, TissueGnostics GmbH, Austria}

\begin{document}
%\ninept
%
\maketitle
\begin{abstract}
Nuclei instance segmentation plays an important role in the analysis of Hematoxylin and Eosin (H\&E)-stained images. While supervised deep learning (DL)-based approaches represent the state-of-the-art in automatic nuclei instance segmentation, annotated datasets are required to train these models. There are two main types of tissue processing protocols, namely formalin-fixed paraffin-embedded samples (FFPE) and frozen tissue samples (FS). Although FFPE-derived H\&E stained tissue sections are the most widely used samples, H\&E staining on frozen sections derived from FS samples is a relevant method in intra-operative surgical sessions as it can be performed fast. Due to differences in the protocols of these two types of samples, the derived images and in particular the nuclei appearance may be different in the acquired whole slide images. Analysis of FS-derived H\&E stained images can be more challenging as rapid preparation, staining, and scanning of FS sections may lead to deterioration in image quality.

In this paper, we introduce CryoNuSeg, the first fully annotated FS-derived cryosectioned and H\&E-stained nuclei instance segmentation dataset. The dataset contains images from 10 human organs that were not exploited in other publicly available datasets, and is provided with three manual mark-ups to allow measuring intra-observer and inter-observer variability. Moreover, we investigate the effects of tissue fixation/embedding protocol (i.e., FS or FFPE) on the automatic nuclei instance segmentation performance of one of the state-of-the-art DL approaches. We also create a baseline segmentation benchmark for the dataset that can be used in future research.  

A step-by-step guide to generate the dataset as well as the full dataset and other detailed information are made available to fellow researchers at \url{https://github.com/masih4/CryoNuSeg}.

\end{abstract}
\section{Introduction}
\label{sec:intro}
Digital pathology enables the acquisition, management and sharing of information retrieved from stained digitised tissue sections from patient-derived biopsies in a digital environment. This offers many benefits including image interpretation by remotely located specialists or further use of the samples for scientific purposes~\cite{doi:10.1111/his.13691}. An additional advantage of digitised samples is their utilisation for computer-mediated quantitative image analysis~\cite{gurcan2009histopathological}. In combination, digital pathology and computational image analysis are expected to significantly improve clinical practice, for instance through second opinion systems supporting the work of pathologists~\cite{barisoni2020digital}.

Examination of Hematoxylin and Eosin (H\&E)-stained tissue sections can reveal important information about individual cells and their functional status~\cite{doi:10.1177/1066896913517939}. Consequently, judgement of these histopathological images remains the ``gold standard'' in diagnosing a variety of diseases including almost all types of cancer. Nuclei morphology, shape, type, count, and density are the key components in the evaluation process of H\&E-stained tissue images. To extract these features automatically with a computer-based method, nuclei instance segmentation is required~\cite{Kumar2017}.

Nuclei instance segmentation masks should provide exact boundaries around each individual nucleus and also distinguish overlapping or touching objects. While many computer-based nuclei instance segmentation methods have been proposed in the literature, supervised deep learning (DL)-based approaches are the most promising~\cite{Kumar2017,monuseg, ZHAO2020101786}. However, they require fully annotated datasets to be able to train the neural networks. Annotated data is also required to quantitatively evaluate any model's performance. While manual labelling of nuclei by biomedical experts is considered as the gold standard method to create nuclei instance segmentation ground truth, this is very time demanding and may suffer from intra-observer and inter-observer variability. %Nevertheless, manually segmented data are essential for the development of fully or semi-automatic supervised computational methods for microscopic image analysis.  

A number of fully manually annotated nuclei instance segmentation datasets of H\&E-stained images have been formerly introduced as shown in Table~\ref{data}. In addition, PanNuke~\cite{gamper2020pannuke} is a recent semi-automatically created dataset of formalin-fixed and paraffin-embedded samples with nearly 200,000 nuclei from 19 different organs where the nuclei masks were first created automatically and then revised by humans. 

\begin{table*}[]
	\caption[]{Publicly available datasets for nuclei instance segmentation. Note that the Kaggle Data Science Dataset contains both H\&E-stained bright-field and fluorescence microscopic images while all others contain only H\&E-stained images.
		
		MoNuSeg = Multi-Organ Nuclei Segmentation; MoNuSAC = Multi-Organ Nuclei Segmentation and Classification; CoNSep = Colorectal Nuclear Segmentation and Phenotypes; CPM: Computational Precision Medicine; TNBC: Triple Negative Breast Cancer; CRCHisto: Colorectaladeno Carcinomas. TCGA: The Cancer Genome Atlas; UHCW: University Hospitals Coventry and Warwickshire.}
	\label{data}
	\begin{tabular}{lcccccc}
		\hline
		\textbf{Dataset} & \textbf{\# image tiles} & \textbf{\# nuclei} & \textbf{magnification} &\textbf{ \# organs} & \textbf{tile size [pixels]} & \textbf{source} \\
		\hline
		Kumar \textit{et al.}~\cite{Kumar2017} & 30 & 21,623 & 40$\times$ & 7 & $1000 \times 1000$ & TCGA                \\
		MoNuSeg~\cite{monuseg} & 44 & 28,846 & 40$\times$ & 9 & $1000 \times 1000$ & TCGA \\
		MoNuSAC~\cite{vermamulti} & 209 & 31,411 & 40$\times$ & 4 & $81 \times 113$ to $1422 \times 2162$ & TCGA \\
		CoNSeP~\cite{2018arXiv181206499G} & 41 & 24,319 & 40$\times$ & 1 & $1000 \times 1000$ & UHCW \\
		CPM-15~\cite{vu2019methods}& 15 & 2,905 & 40$\times$, 20$\times$ & 2 & $400 \times 400$, $600 \times 1000$ & TCGA \\
		CPM-17~\cite{vu2019methods} & 32 & 7,570 & 40$\times$, 20$\times$ &4& $500 \times 500$ to $600 \times 600$ & TCGA \\
		TNBC~\cite{naylor2018segmentation} & 50 & 4,022 & 40$\times$ &1 & $512 \times 512$ & Curie Inst. \\
		CRCHisto~\cite{sirinukunwattana2016locality} & 100 & 29,756 & 20$\times$ &1 & $500 \times 500$ & UHCW \\
		Kaggle Data Science~\cite{caicedo2019nucleus} & 670 & 29,464 & n/a & n/a & $256 \times 256$ to $1040 \times 1388$ & n/a \\
		Janowczyk~\cite{Janowczyk_website} & 143 & 12,000 & 40$\times$ &1 & $2000 \times 2000$ & n/a \\
		Crowedsource~\cite{irshad2014crowdsourcing} & 64 & 2,532 & 40$\times$ &1 & $400 \times 400$ & TCGA \\
		\hline
	\end{tabular}
\end{table*}   

\begin{figure*}
	\centering
	\includegraphics[width=16cm,height=4.56cm]{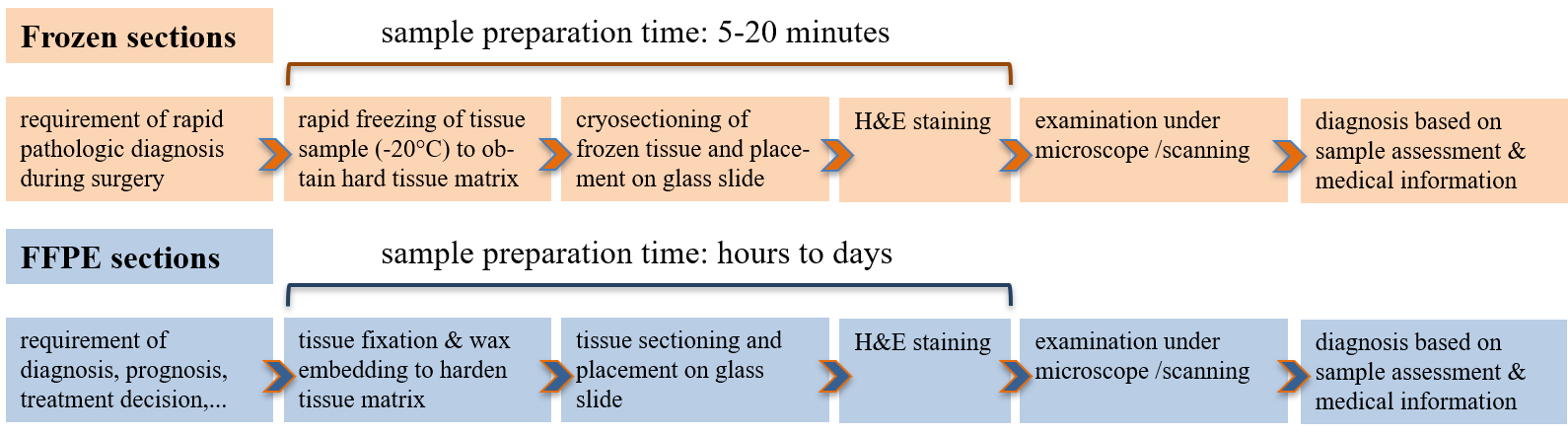} 
	\caption{Preparation workflow of frozen sections (top) and formalin-fixed paraffin-embedded (FFPE) sections (bottom). The fixation and embedding procedure of FFPE samples is much more time consuming compared to the freezing step required to prepare frozen tissue samples.}
	\label{dif}
\end{figure*}

While these datasets contain H\&E-stained images and corresponding instance segmentation ground truths, an important information regarding the utilised tissue processing approach is missing. There are two main types of tissue processing methods that can be applied before H\&E-staining of sections, namely formalin (chemical)-fixation and paraffin-embedding (FFPE) of tissue samples, and preparation of frozen samples (FS). The general workflows to obtain H\&E-stained sections using these two methods are shown in Fig.~\ref{dif}. 

H\&E staining of FFPE samples is the more widely used approach. Here, samples are fixed with formalin and embedded in blocks of paraffin wax. Thin slices of these blocks are then stained and used for diagnostic purposes. While the samples derived form this workflow are very durable, the entire procedure is rather time-consuming, taking hours to days~\cite{PMID:9199619}. On the other hand, preparation of sections from FS is very fast and takes between 5 and 20 minutes~\cite{PMID:9199619}. The samples are rapidly frozen (for structure preservation)  and then a cryo-microtome is used to section them.

Following H\&E-staining, the samples are investigated under a microscope. FS staining is usually required in oncological surgeries, where rapid and online diagnosis and decisions are important for further processes. Distinguishing a benign lesion from a malignant lesion, online cancer grading, determining the borders around the abnormal tissues, and accessing the invasion status of lesions are some applications where FS are used~\cite{jaafar2006intra}. A 92\%--98\% diagnostic accuracy with images derived from FS can be achieved by experienced pathologists. However, in some challenging cases, different diagnostic outcomes are reported for FFPE- and FS-derived H\&E stained images. This is mainly due to technical problems of rapid sample preparation in FS-derived samples which may lead to poor quality of the derived whole slide images (WSIs). Due to the same technical issue, the nuclei appearance may be different in FS-derived samples compared to FFPE-derived samples. More condensed nuclear chromatin or nuclear ice crystals are some of the artefacts of FS-derived stained sections~\cite{jaafar2006intra}. These artefacts can not only affect the interpretation by medical experts but also may have a significant impact on the performance of DL-based algorithms for nuclei instance segmentation. In general, due to insufficient numbers of available FS-derived H\&E-stained WSIs, only few studies have analysed FS-derived H\&E-stained tissue sections~\cite{komura2018machine, 10.1117/12.2542799, 10.1117/12.2538303, 10.1117/12.2216448} and, to our knowledge, no research has been conducted to explicitly analyse nuclei instance segmentation of FS-derived H\&E-stained images.

In this paper, we present CryoNuSeg, the first fully annotated nuclei instance segmentation dataset based solely on FS-derived H\&E-stained images. Nuclei are labelled manually by two experts, which thus allows the measurement of inter-observer variability. Moreover, one of the annotators re-labelled the entire dataset so that intra-observer variability can also be investigated. The dataset, which is available on the Kaggle platform\footnote{\url{https://www.kaggle.com/ipateam/segmentation-of-nuclei-in-cryosectioned-he-images}}, comprises images from human organs that have not been used in formerly released datasets, and we provide step-by-step descriptions of sample selection, sample preparation, and generation of segmentation masks throughout this paper and in the publicly available repository at \url{https://github.com/masih4/CryoNuSeg}. We further exploit a state-of-the-art DL-based nuclei segmentation algorithm~\cite{10.1007/978-3-030-23937-4_9} and investigate the effect of tissue fixation/embedding protocol (i.e., FS or FFPE) on instance segmentation performance. We also create a baseline segmentation benchmark for the dataset that can be used in future studies.

\section{Method}
\label{sec:format}

\subsection{Dataset}
\label{dataset}
The Cancer Genome Atlas (TCGA)\footnote{\url{https://portal.gdc.cancer.gov/repository}} contains more than 30,000 WSIs from more than 50 human organs and tissues. Using the filtering options provided in TCGA, we selected FS-derived H\&E-stained images acquired at 40x magnification and chose images from organs not present in other publicly available datasets. With the help of a senior cell biologist from the Medical University of Vienna, we selected 30 WSIs from 10 different human organs (three WSIs per organ), namely the adrenal gland, larynx, lymph node, mediastinum, pancreas, pleura, skin, testis, thymus, and thyroid gland. To maximise data variability, we also aimed to include samples from different scanning centres, different disease types and different sexes. Using QuPath\footnote{\url{https://qupath.github.io/}}, we extracted image patches of a fixed size of $512 \times 512$ pixels (one image patch per WSI) while aiming to extract the patches from the most representative parts of the WSIs. To perform manual segmentation, we used ImageJ\footnote{\url{https://imagej.nih.gov/ij/}} and its pre-built region of interest (ROI) manager tool. Further technical step-by-step descriptions of WSI selection, WSI patch extraction by QuPath, and manual annotation with ImageJ are detailed in the released GitHub repository. Manual instance nuclei segmentation was performed by two trained annotators, a biologist (Annotator 1) and a bioinformatician (Annotator 2). Both annotators were instructed in the same way to segment the nuclei.  Annotator 1 also performed a second round of manual mark-ups with a gap of about three months between the two annotations.

In the entire dataset, Annotator 1 identified and segmented 7,596 and 8,044 nuclei in the first and second round of manual mark-ups, respectively while Annotator 2 segmented 8,251  nuclei. %We used Matlab (version 2018a) to convert the ImageJ ROI files to conventional binary and labelled masks.
Besides creating labelled and binary masks, we also created additional auxiliary segmentation masks that can be useful in training supervised DL-based approaches. These auxiliary masks include binary segmentation masks generated by removing touching borders, distance maps, and weighted maps that give more weight to the pixels between close nuclei. Such masks have been shown useful for nuclei instance segmentation in former studies~\cite{Ronneberger2015, 10.1007/978-3-030-23937-4_9, naylor2018segmentation, vu2019methods}, and the related codes to create both conventional and auxiliary segmentation masks are made available in our GitHub repository.

\begin{figure*}
	\centering
	\includegraphics[width=.9\textwidth]{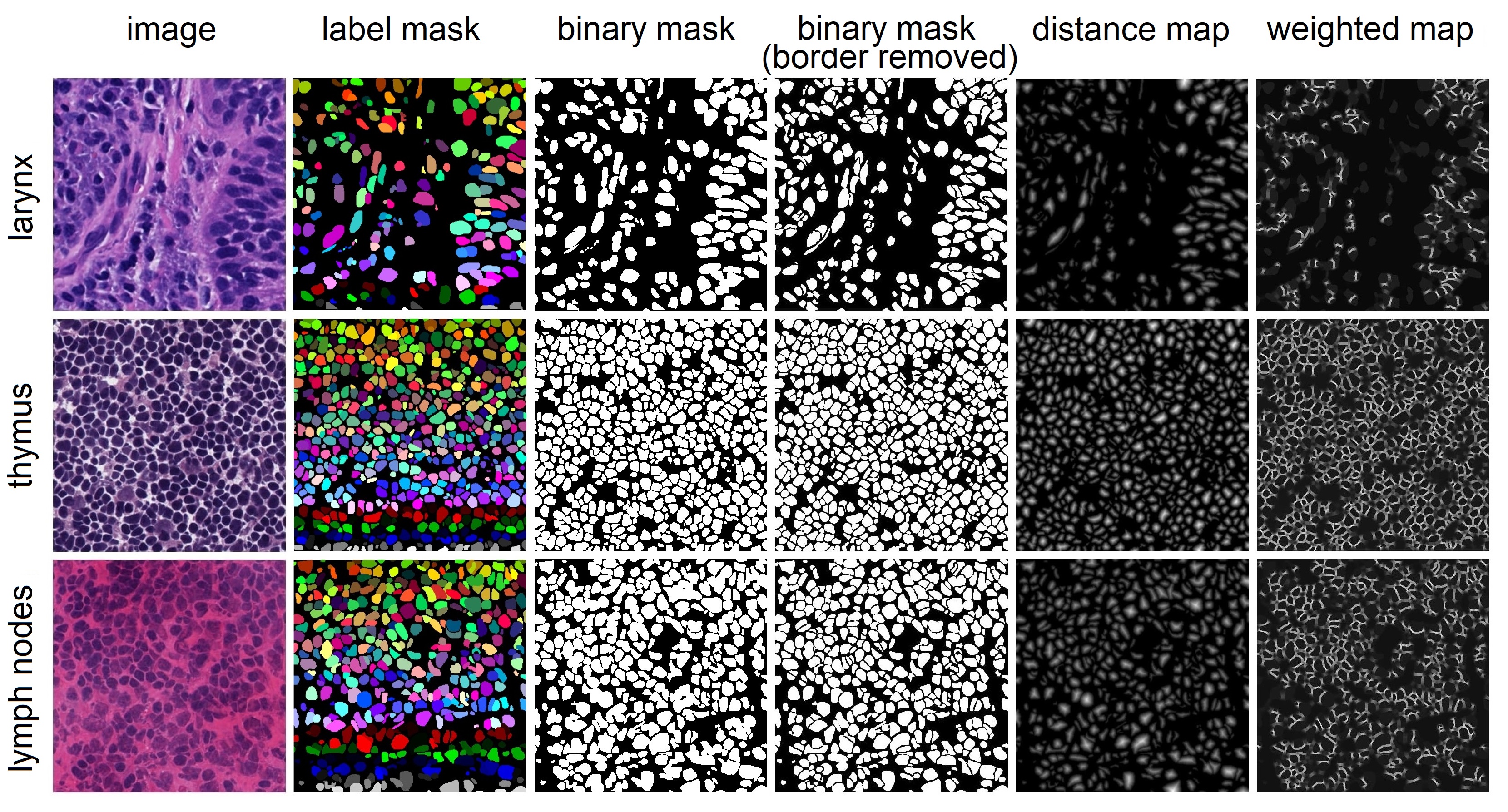}
	\caption{Sample CryoNuSeg images from three human organs and their corresponding segmentation masks (from first segmentations of Annotator 1). For each sample, we show (from left to right) the raw image patch, the manually labelled nuclei, the binary segmentation mask, the binary mask with touching borders removed, the distance map, and the weighted map.}
	\label{examples}
\end{figure*}

\begin{figure}
	\centering
	\includegraphics[width=\columnwidth]{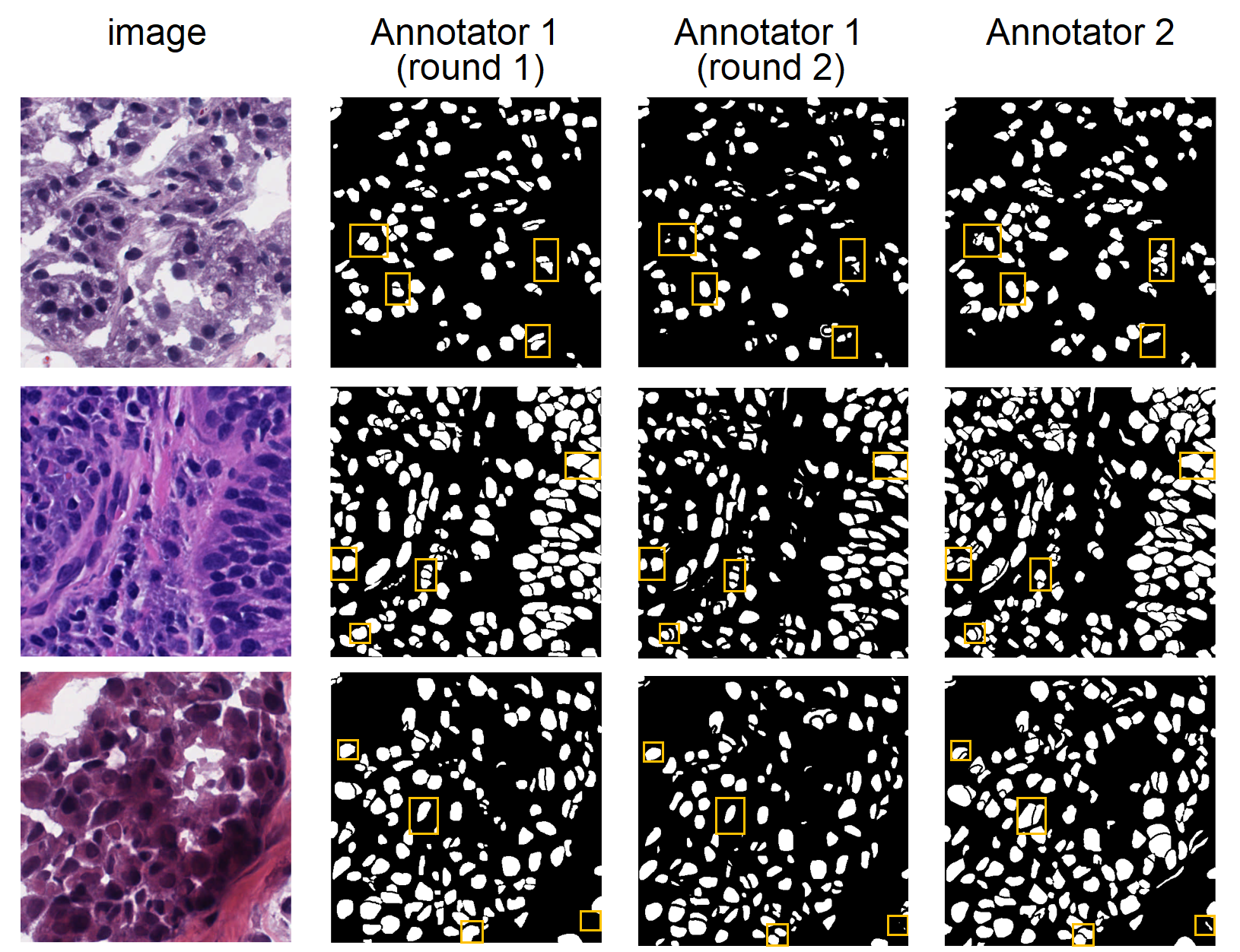}
	\caption{Examples of deviations in the annotations. Some mismatching annotations exemplifying inter-observer variability (comparison between second/third column and fourth column) and intra-observer variability (comparison between second and third column) are shown in yellow boxes. While there are more mismatched pairs in the masks, we limit the number of mismatched pairs to four for better visualisation.}
	\label{vary}
\end{figure}

Fig.~\ref{examples} gives some examples of raw image patches together with the resulting conventional and auxiliary segmentation masks. In Fig.~\ref{vary}, we show some inconsistent cases between the segmentation masks of Annotator 1 and Annotator 2 (inter-observer variability) and some mismatched cases between the first and second segmentations of Annotator 1 (intra-observer variability). %The full CryoNuSeg dataset that we release includes the raw image patches, the raw ROI ImageJ files, and derived segmentation masks.

To investigate the effects of the tissue fixation/embedding protocol on DL-based nuclei segmentation, we use the MoNuSeg dataset~\cite{monuseg} mentioned in Table~\ref{data}. Information regarding the tissue processing type is not provided in former datasets, however, since TCGA is also the data source of the MoNuSeg dataset, by tracking down the image codes it is possible to retrieve the tissue fixation/embedding protocol (FS or FFPE) of the MoNuSeg images. Of the 44 image patches in the MoNuSeg dataset, 27 images are from FS-derived sections and 17 from FFPE-derived sections. Based on this, we separate the images to form two subsets, namely MoNuSeg-FS and MoNuSeg-FFPE. We use the same number of images per organ to be able to compare the nuclei segmentation results. For instance, of the nine kidney images in MoNuSeg, seven are FS and two FFPE samples. Thus, to have a balanced datasets, we randomly choose two FS kidney images for the MoNuSeg-FS subset. We exclude colon and lung images since their samples are all of the same tissue fixation/embedding protocol. The number of annotated nuclei in the derived MoNuSeg-FS and MoNuSeg-FFPE datasets are 7639 and 7503, respectively, and are split amongst the organs as shown in Table~\ref{two_datasets}.  

\begin{table}[]
	\caption[]{Number of images and nuclei per organ in the MoNuSeg-FS and MoNuSeg-FFPE datasets.}
	\label{two_datasets}
	\begin{tabular}{l|cc|cc}
		\hline
		\multirow{2}{*}{organ} & \multicolumn{2}{c|}{MoNuSeg-FS} & \multicolumn{2}{c}{MoNuSeg-FFPE} \\ \cline{2-5} 
		& \# images       & \# nuclei   & \# images   & \# nuclei  \\
		\hline
		%lung            &  0          &   0       &  0       &  0               \\
		%colon            &  0          &   0       &  0       &  0               \\
		breast           &  3          &  1,267     &  3       &  1,117               \\
		bladder          &  1          &   677     &  1       &  342               \\
		prostate         &  4          &  1,716     &  4       &  1,475               \\
		brain            &  1          &  465      &  1       &  249              \\ 
		stomach          &  1          &   1,165    &  1       &  1,390               \\ 
		liver            &  3          &   1,461    &  3       &  1,282               \\ 
		kidney           &  2          &   888     &  2       &  1,648               \\ 
		\hline
		total            &  15         &  7,639     &  15      &  7,503              \\ 
		\hline
	\end{tabular}
\end{table}

\subsection{Nuclei segmentation method}
\label{model}
\begin{figure*}
	\centering
	\includegraphics[width=.7\textwidth]{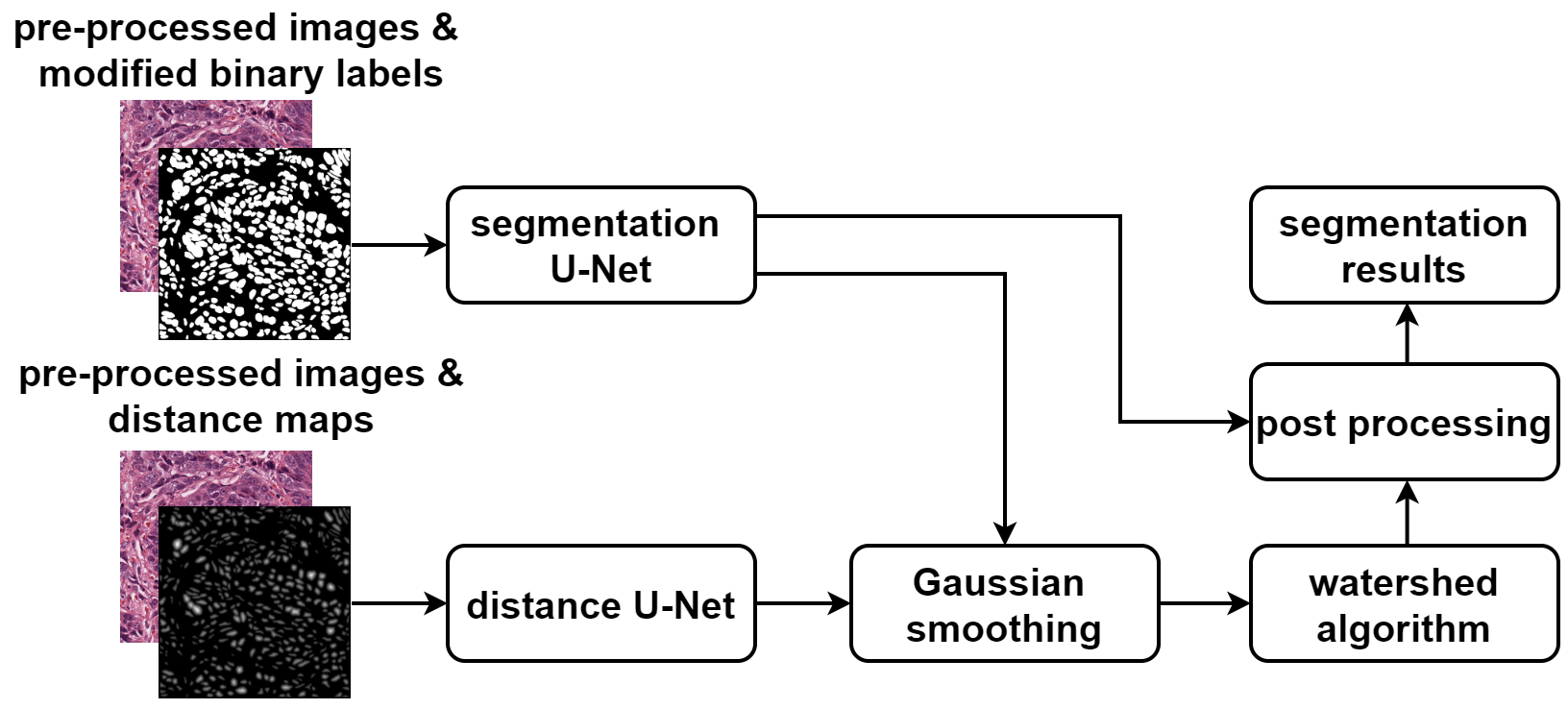}
	\caption{Flowchart of the employed instance segmentation algorithm.}
	\label{seg_model}
\end{figure*}

We use our recently published state-of-the-art DL-based instance segmentation algorithm~\cite{10.1007/978-3-030-23937-4_9} as the baseline segmentation model. The general workflow of this method is illustrated in Fig.~\ref{seg_model}. 

The method proceeds in two stages. A U-Net~\cite{Ronneberger2015}-based model (segmentation U-Net) is used to separate foreground and background, while a regression encoder-decoder-based model (distance U-Net) is employed to predict the distance maps for all nuclei instances. We apply a Gaussian smoothing filter on the predicted distance maps to prevent false local maxima detection, with the kernel size of the filter determined from the segmentation U-Net results based on the average predicted nuclei size. The results from the two stages are then merged using a watershed-based algorithm. Each of the sub-models is trained for 30 epochs with a learning rate scheduler (the starting learning rate of 0.01 is halved after every 8 epochs). We set the batch size to 8 and apply various augmentation techniques in the training phase including random horizontal and vertical flipping (with a probability of 0.5), random 90, 180, and 270 degrees rotations (with a probability of 0.5), random brightness and contrast shift (brightness and contract shift limit of 0.15 with a probability of 0.4), random adaptive histogram equalisation (tile grid size of 8 with the probability of 0.5) and random image cropping with a fixed size of 512 $\times$ 512 pixels~\cite{MAHBOD2020105475, MAHBOD2020105725}. In the inference phase, we apply two post-processing steps on the final segmentation results which remove very small instances from the segmentation masks (objects with areas containing less than 20 pixels) and fill holes inside detected objects (using morphological operations).

\subsection{Evaluation}
To evaluate the segmentation performance, we employ three evaluation indexes~\cite{Kirillov_2019_CVPR, Kumar2017, 2018arXiv181206499G}, namely the Dice score, aggregate Jaccard index (AJI), and panoptic quality (PQ) score. The Dice score evaluates the general performance of semantic segmentation, while both AJI and PQ measure instance segmentation performance. Compared to AJI, PQ is a more robust score and does not suffer from over-penalisation. To detect statistically significant differences in the obtained results, we use a two-sided Wilcoxon signed-rank test~\cite{gibbons2014nonparametric}.

\subsection{Implementation}
\label{ssec:subhead}
The code to generate the CryoNuSeg segmentation masks from the ImageJ ROI files is implemented in Matlab 2018a, while for the segmentation model we use the Keras DL framework (version 2.3.1). All experiments are performed on a single workstation with an Intel Core i7-8700 3.20 GHz CPU, 32 GB of RAM, and a TITIAN V NVIDIA GPU card with 12 GB of installed memory.

\section{Results}
\label{sec:print}
We first use the MoNuSeg-FS dataset as well as the MoNuSeg-FFPE dataset as training sets and employ the entire CryoNuSeg dataset as the test set. We perform an identical training scheme for training both models as described in Section~\ref{model}. The main aim of this set of experiments is to investigate the effect of tissue fixation/embedding protocol on the performance of the nuclei segmentation model. The obtained results are given in Table~\ref{results1}.

\begin{table*}[]
	\caption{Segmentation results on CroNuSeg when trained on MoNuSeg-FS and MoNuSeg-FFPE, respectively. The results, in terms of Dice score, AJI, and PQ score, are based on the manual segmentation masks from Annotator 1 (first round of mark-ups) as ground truth. The reported $p$-values in the bottom row show the results of the statistical significance test when comparing the two training sets.}
	\label{results1}
	\begin{tabular}{l|cc|cc|cc}
		\hline
		\multirow{2}{*}{\textbf{test organ}} & \multicolumn{2}{c|}{\textbf{Dice score (\%)}} & \multicolumn{2}{c|}{\textbf{AJI (\%)}} & \multicolumn{2}{c}{\textbf{PQ score (\%)}} \\ \cline{2-7} 
		& \small \textbf{MoNuSeg-FS}         & \small \textbf{MoNuSeg-FFPE}         & \small \textbf{MoNuSeg-FS}         & \small \textbf{MoNuSeg-FFPE}        & \small \textbf{MoNuSeg-FS}        & \small \textbf{MoNuSeg-FFPE}        \\
		\hline
		\textbf{adrenal gland}      & 78.31     & 80.80      & 51.64           & 57.11       & 47.01          &  52.08           \\
		\textbf{larynx}             & 80.54     & 80.80      & 57.61           & 57.68       & 54.56          &  54.65           \\
		\textbf{lymph node}        & 79.71     & 80.00      & 49.62           & 49.90       & 49.07          &  49.18           \\
		\textbf{mediastinum}        & 83.24     & 83.34      &  50.56          & 50.84       & 47.19          &  47.85           \\
		\textbf{pancreas}           & 67.43     & 66.06      &  38.04          & 37.18       & 32.43          &  31.91           \\
		\textbf{pleura}             & 71.74     & 69.20      &  42.96          & 41.74       &  37.31         &  36.22           \\
		\textbf{skin}               & 72.96     & 72.33      &  45.82          & 45.73       &  40.29         &  40.40           \\
		\textbf{testis}             & 81.08     & 82.20      &  48.54          & 50.21       & 44.20          &  44.96           \\
		\textbf{thymus}             & 84.50     & 84.66      &   53.36         & 53.76       & 48.35          &  48.97           \\
		\textbf{thyroid gland}      & 80.11     & 80.46      &  55.80          & 56.83       &  48.19         &  51.62           \\ \hline
		\textbf{average}            & 78.0 $\pm$ 5.5    & 78.0  $\pm$ 6.4     &  49.4 $\pm$ 5.9   &  50.1  $\pm$ 6.8     &  44.9 $\pm$ 6.5 &  45.8 $\pm$  7.4   \\ 
		\hline
		\textbf{$p$-value}     &\multicolumn{2}{c|}{0.6800} &\multicolumn{2}{c|}{0.1529} &\multicolumn{2}{c}{0.0719}\\ 
		\hline
	\end{tabular}
\end{table*}

We perform an additional experiment using the combined MoNuSeg-FS and  MoNuSeg-FFPE datasets as training data and the entire CryoNuSeg dataset as test set. The results of this yield an average Dice score of 79.1 $\pm$ 5.2\%, average AJI of 51.8 $\pm$ 6.4\%, and average PQ score of 48.2 $\pm$ 7.3\%. A statistical test comparing the results derived from the combined MoNuSeg-FS/MoNuSeg-FFPE dataset gives $p$-values of 0.0196, 0.0001, and 0.0001 for Dice score, AJI, and PQ score, respectively, when comparing to the results from MoNuSeg-FS, while $p$-values of 0.0068 (Dice), 0.0005 (AJI), and 0.0001 (PQ) are obtained when comparing to the results derived from the MoNuSeg-FFPE dataset.

In the next experiment, we use the same segmentation model but perform 10-fold cross-validation (10CV) on the CryoNuSeg dataset to obtain segmentation results. In each fold, 27 images from nine organs are used for training, and the trained model is then evaluated on the hold-out test organ. The motivation behind this experiment is to create a baseline segmentation benchmark that should prove useful for future studies based on the introduced CryoNuSeg dataset. The results of this experiment are shown in Table~\ref{results2}.

\begin{table}[]
	\caption{10CV segmentation results of CryoNuSeg dataset in terms of Dice score, AJI, and PQ score (based on the segmentation masks of Annotator 1 in the first round of mark-ups).}
	\label{results2}
	\begin{tabular}{l|ccc}
		\hline
		\textbf{test organ} & \textbf{Dice (\%)} & \textbf{AJI (\%)} & \textbf{PQ (\%)} \\ \hline
		\textbf{adrenal gland}  &  78.17         & 53.49    & 48.30   \\
		\textbf{larynx}         &   81.65        & 59.70     & 54.50   \\
		\textbf{lymph node}    &   81.56        & 53.54    & 50.79   \\
		\textbf{mediastinum}    &   84.87        & 54.10    & 50.73   \\
		\textbf{pancreas}       &   74.31        & 44.84    &  37.75  \\
		\textbf{pleura}         &   75.49        & 46.49    &  40.02  \\
		\textbf{skin}           &   74.75        & 47.84    &  40.78  \\
		\textbf{testis}         &   84.27        & 50.49    &  47.51  \\
		\textbf{thymus}         &   85.90        & 56.46    & 52.83   \\
		\textbf{thyroid gland}  &   81.81        & 58.20   &  53.48  \\ 
		\hline
		\textbf{average}        & 80.3 $\pm$ 4.3  &  52.5 $\pm$ 5.0   &  47.7 $\pm$ 6.1  \\ 
		\hline
	\end{tabular}
\end{table}

We perform an additional 10CV experiment where in each fold 27 images from nine organs of the CryoNuSeg dataset together with the combined MoNuSeg-FS/MoNuSeg-FFPE dataset is used for training and the hold-out organ from the CryoNuSeg dataset is used as the test set (i.e., there are 57 training images and 3 test images in each fold). This results for this are an average Dice score of  80.2 $\pm$ 4.8\%, an average AJI of 52.9 $\pm$ 5.7\%, and an average PQ score of 49.4 $\pm$ 6.5\%.

While in the former experiments, we choose the segmentation masks from Annotator 1 (first round of manual mark-ups) as the ground truth, in our next experiments, we investigate the inter-observer variability between the two annotators. For this, we compare the manual segmentation masks generated by Annotator 2 to those from Annotator 1 (first round) with the latter being used as ground truth. The results are reported in Table~\ref{results3}.

\begin{table}[]
	\caption{Segmentation results, in terms of Dice score, AJI, and PQ score, from comparing the manual segmentation masks from Annotators 2 to those from Annotator 1 (first round) to show inter-observer variability.}
	\label{results3}
	\begin{tabular}{l|ccc}
		\hline
		\textbf{test organ} & \textbf{Dice (\%)} & \textbf{AJI (\%)} & \textbf{PQ (\%)} \\ \hline
		\textbf{adrenal gland}  &  78.08         & 53.16   & 49.77   \\
		\textbf{larynx}         &  83.68         & 62.53   & 58.39   \\
		\textbf{lymph node}    &  82.38         & 57.80   &  54.80  \\
		\textbf{mediastinum}    &  84.09         & 58.34   & 54.36   \\
		\textbf{pancreas}       &  67.14         & 40.66   & 35.93   \\
		\textbf{pleura}         &  69.74         & 45.09   &  42.62  \\
		\textbf{skin}           &  74.06         & 47.89   & 46.55   \\
		\textbf{testis}         &  83.86         & 55.37   & 51.12   \\
		\textbf{thymus}         &  85.88         & 62.00   & 58.49   \\
		\textbf{thyroid gland}  &  80.52         & 58.36   & 56.57   \\ 
		\hline
		\textbf{average}        &  78.9 $\pm$ 6.5 & 54.1 $\pm$ 7.3   &  50.9 $\pm$ 7.3  \\ 
		\hline
	\end{tabular}
\end{table}

To measure intra-observer variability, we compare the first and second segmentation masks of Annotator 1 and show the results in Table~\ref{results4}.

\begin{table}[]
	\caption{Segmentation results, in terms of Dice score, AJI, and PQ score, from comparing Annotator 1's manual segmentation masks of the first round to those of the second round to show intra-observer variability.}
	\label{results4}
	\begin{tabular}{l|ccc}
		\hline
		\textbf{test organ} & \textbf{Dice (\%)} & \textbf{AJI (\%)} & \textbf{PQ (\%)} \\ \hline
		\textbf{adrenal gland} & 84.83   & 62.25      & 55.44   \\
		\textbf{larynx}        & 84.92   & 67.25      & 63.63   \\
		\textbf{lymph node}   & 84.77   & 63.22      & 59.46   \\
		\textbf{mediastinum}   & 84.67   & 62.40      & 57.32  \\
		\textbf{pancreas}      & 78.27   & 54.63      & 50.71   \\
		\textbf{pleura}        & 82.17   & 61.00      & 53.37  \\
		\textbf{skin}          & 78.95   & 55.16      & 53.69  \\
		\textbf{testis}        & 86.21   & 61.42      & 54.12  \\
		\textbf{thymus}        & 87.84   & 67.77      & 60.90  \\
		\textbf{thyroid gland} & 85.69   & 68.79      & 61.66  \\ 
		\hline
		\textbf{average}        &   83.8$\pm$3.1  &  62.4$\pm$ 4.8    &   57.0$\pm$4.2   \\ 
		\hline
	\end{tabular}
\end{table}

\section{Discussion}
In this paper, we introduce the first fully manually annotated nuclei segmentation dataset based on FS-derived H\&E-stained sections. % from 10 human organs that were not presented in former studies. 
Further, we investigate the impact of the tissue fixation/embedding procedure on the segmentation performance of a state-of-the-art DL-based nuclei segmentation algorithm. Our dataset can be used alongside other publicly available datasets to train supervised machine learning-based approaches or can be used as a stand-alone benchmark to evaluate the performance of nuclei segmentation methods.  

The description of the dataset, the steps to generate it, and the related implementation codes to create conventional and auxiliary segmentation masks from the ImageJ ROI files are available in the published Github repository, while the dataset itself is available on the Kaggle website. With Kaggle's cloud-based computational resources, it is possible to use the dataset and develop instance segmentation models directly on the Kaggle website. We have implemented an example kernel with the Jupyter notebook on the Kaggle platform based on the well-known U-Net algorithm with watershed post-processing~\cite{Ronneberger2015, Yang2006}. This kernel should prove helpful to show how to use the dataset on the Kaggle platform and is publicly available on the dataset page under the notebook section\footnote{\url{https://www.kaggle.com/ipateam/u-net-with-binary-labels}}.  

The results in Table~\ref{results1} show the impact of the tissue fixation/embedding procedure on instance segmentation performance based on training models derived from the MoNuSeg-FS and MoNuSeg-FFPE datasets which contain only FS-derived and only FFPE-derived H\&E-stained images, respectively. While FS- and FPPE-derived images can never be obtained from the exact same sample, we have tried to minimise all other data variability differences using the same number of images per organ and samples of the same organs (breast, bladder, prostate, brain stomach, liver, and kidney) which results also in roughly the same number of segmented nuclei (7639 for MoNuSeg-FS and 7503 for MoNuSeg-FFPE). As Table~\ref{results1} shows, the overall and organ-wise segmentation results are very competitive (except for AJI and PQ score for the adrenal gland) in both cases (i.e., when trained on MoNuSeg-FS and MoNuSeg-FFPE, respectively). The Wilcoxon signed-rank test yields $p$-values larger than 0.05 for all three evaluation measures (Dice score, AJI, and PQ score). Thus, we find nuclei instance segmentation performance to not be significantly affected by the employed tissue fixation/embedding procedure. As there are rather large segmentation performance differences based on AJI and PQ score for the adrenal gland for the MoNuSeg-FS and MoNuSeg-FFPE datasets, further research with a larger training sample of adrenal gland images is required to investigate the impact of tissue fixation protocol on the nuclei instance segmentation for this type of tissue.  

The results from the combined MoNuSeg-FS/MoNuSeg-FFPE experiment show a slight but statistically significant improvement in the segmentation performance, specifically for PQ score, with the boost in segmentation performance likely related to the increased number of training images (30 instead of 15 images) in this experiment. Further research is required to investigate the impact of the tissue fixation/embedding protocol on other histological image analysis tasks such as gland/tumour segmentation or WSI classification/grading.

As a segmentation benchmark for our CryoNuSeg dataset, we perform 10-fold cross-validation whereby the images of nine organs are used for training while testing on the tenth and repeating the process so that each organ is once used for testing. This separation method had two advantages over a random division. First, the segmentation results reported in Table~\ref{results2} show the generalisation ability of the model better, since in each fold an unseen test organ is used for the evaluation. Second, for future studies that use this dataset for developing instance segmentation models, identical folds can be easily created to fairly compare segmentation results.

The results from the experiment, where MoNuSeg-FS and MoNuSeg-FFPE are also used for training, show, as expected, a slight improvement in the segmentation performance.

Comparing the results from Table~\ref{results1} and Table~\ref{results2}, we can see that segmentation performance improves when training on CryoNuSeg images compared to training on MoNuSeg images (either MoNuSeg-FS or MoNuSeg-FFPE). The difference is statistically significant as confirmed by a pair-wise statistical test for each of the utilised evaluation indices. The obtained $p$-values are well below 0.05 %0.003 
for all six cases (three evaluation measures each for MoNuSeg-FS vs.\ CryoNuSeg and MoNuSeg-FFPE vs.\ CryoNuSeg). This could be related to the larger number of training images in the CryoNuSeg dataset compared to the MoNuSeg-FS/MoNuSeg-FFPE images although all three datasets have roughly the same number of annotated nuclei. Another reason could be related to the inter-observer variability since CryoNuSeg, a single person annotated the training and test images (Annotator 1 in our experiments), while for the MoNuSeg-FS and MoNuSeg-FFPE images there are different annotators for the training and test images. 
% it's worth mentioning but not worth mentioning again
%It is worth mentioning that besides the last experiment in this paper (i.e. results in Table~\ref{results3}), we only used the manual segmentation masks from the first annotator (biologist) in all experiments related to the CryoNuSeg dataset.  

Our CryoNuSeg dataset comes with segmentations from two annotators (a biologist and a bioinformatician); both were instructed identically on how the segmentation task should be obtained (from a technical viewpoint). To measure the inter-observer variability, Table~\ref{results3} compares the segmentation masks of two annotators. While in an ideal scenario, a perfect match would be achieved, as the results show there is a significant difference. The difference is relatively small in terms of Dice score but more evident for AJI and PQ score. This indicates that the overall agreement of the two annotators to distinguish the background and foreground is much better than to distinguishing touching or overlapping nuclei. By visual inspection of the manual segmentation masks from the two annotators, it can be noticed that Annotator 2 has a tendency to over-segment the nuclei as shown by some examples in Fig.~\ref{vary} and as demonstrated by the higher number of identified nuclei compared to Annotator 1. The issue of inter-observer variability was also observed in a subset of the MoNuSeg dataset where the agreement between two annotators (based on AJI) was reported to be only 65\%~\cite{monuseg}. 

Since two segmentations are available from Annotator 1, we can measure intra-observer variability as reported in Table~\ref{results4}. While a perfect match would represent the ideal situation, as the results in Table~\ref{results4} show, there is a significant difference for all three evaluation indexes (and much more evident for AJI and PQ score). By comparing the average and organ-wise results in Table~\ref{results3} and Table~\ref{results4}, it can be observed that inter-observer variability has a larger impact on the segmentation performance in comparison to intra-observer variability.  

Fuzzy or unclear borders between touching nuclei, folded tissues and other acquisition artefacts, manual annotation errors in the nuclei borders, and sensitivity loss of the annotators due to fatigue are some parameters that can cause inter-observer and intra-observer variability issues. These problems can be partially resolved by removing vague areas in the manual segmentation masks~\cite{vermamulti}, but this requires extra supervision and time.

We can also compare the manual segmentation results of the second annotator from Table~\ref{results3} with the 10CV results from Table~\ref{results2} to fairly compare an automated segmentation algorithm with a human annotator. From the tables, we can see that the results are relatively close. Statistical tests for the three evaluation indexes yield $p$-values of 0.055 for Dice score, 0.052 for AJI, and 0.003 for PQ score which suggest that the baseline segmentation method we provide is within reach of the performance of a manual annotator (although the PQ score results are statistically significantly different). %Another interesting topic that can be addressed in future studies is further measuring the intra-observer variability and compare it to the automatic segmentation results.        

\section{Conclusions}
In this paper, we have introduced the first fully manually annotated FS-derived H\&E-stained nuclei segmentation dataset. Our CryoNuSeg dataset is publicly available to fellow researchers together with extensive documentation and sample implementations. It can thus be used for developing and comparing nuclei instance segmentation algorithms. We have also provided a baseline segmentation benchmark founded on a state-of-the-art DL-based approach for this purpose. In addition, we have investigated the impact of the tissue fixation/embedding method on the segmentation performance, while further work is planned to investigate and compare the impact of tissue fixation and embedding technique on different histological image analysis tasks. We hope that the dataset will prove useful for the community and are looking forward to its use in future studies.

\section*{Acknowledgements}
\label{sec:acknowledgments}
This work was supported by the Austrian Research Promotion Agency (FFG), No. 872636, and a Kaggle open data research grant. We would like to also thank NVIDIA for their generous GPU donation.

% -------------------------------------------------------------------------
\bibliographystyle{IEEEbib}
\bibliography{CBM2020}

\end{document}